# Social Network Conceptualization and Operationalization of Hierarchy Within Therapeutic Communities

*Benjamin W. Campbell, The Ohio State University*
*Keith Warren, The Ohio State University*

## Introduction

In the United States, the Therapeutic Community (TC) treatment model for substance use disorder recovery relies heavily upon the assumption that peers serve as mentors, affirming behavior consistent with the norms of the treatment model and correcting behavior contrary to those norms. Despite being foundational to the TC model, little work has been done to clearly conceptualize and operationalize this hierarchy. In this manuscript, we fill that gap in the literature, presenting a novel conceptualization of hierarchy within the TC context, complete with a complementary measurement drawing from the field of social network analysis. With this new framework for understanding and studying hierarchy within TCs, we can ask and rigorously answer a new set of questions previously unexamined.

Upon presenting our new framework for studying hierarchy, we ask and answer three questions to generate novel insights and use data to assess conventional clinical assumptions. First, we examine the "shape" of the hierarchy which allows us to get a sense of the overall pattern of mentorship relationships and whether the hierarchy is bottom-heavy (triangle), top-heavy (inverted triangle), or somewhere in the middle (diamond). Ultimately, we find that it is the latter, with a handful at the top, many in the middle, and few in the bottom.

Second, this framework allows us determine whether one's position in the hierarchy is heavily associated with their seniority. If we can confirm this clinical wisdom with a rigorous data analysis, it is good news for the TC model as it would imply that one acts as a good member and internalizes the community norms as they spend more time in the facility. Alternatively, if we find countervailing evidence, the TC model would be problematized as we would find that those who are the most active and influential peers might not necessarily be those who have the most time working through the treatment program. We find good news, that one's terminal position in the hierarchy is on average higher than their initial position in the hierarchy.

Finally, we examine the critical issue of whether one's terminal position in the hierarchy predicts their successful completion and graduation of their treatment program. Similar to the prior question, it would be troubling if those who were at the higher terminal positions in the hierarchy, mentoring and advising others in the TC, were not also more likely to successfully complete their treatment program than those with lower terminal positions in the hierarchy. Indeed, clinicians would be worried if the most active mentors and those at the highest positions in the hierarchy were themselves unlikely to graduate from their treatment program. Conversely, clinicians would be worried if the residents who they would want to be mentoring others were more closed off and inactive. Again, we find good news, the higher one's terminal position in the hierarchy, the greater their likelihood of successfully completing their treatment program.

Before we get to these novel findings, however, we will start with a literature review and then proceed to our framework for considering hierarchy.

## Literature Review

Relative to the importance of hierarchy to the TC treatment model of substance use disorder recovery, conceptualizing and operationalizing hierarchy has long eluded researchers. When discussing TCs, it is important to note that the tradition of British TCs differ from those of American TCs with respect to the importance of hierarchy. British TCs, developed by the psychiatrist Maxwell Jones, were used following World War II as residential treatment programs for soldiers we may contemporarily diagnose with Post Traumatic Stress Disorder (Vandevelde et al, 2004; Pearce & Haigh, 2017). In the original British application, clinicians sought a radical flattening of the hierarchy, wherein residents and staff were largely encouraged to reflect on their own behavior (Pearce & Haigh, 2017). These "democratic" TCs were later extended to treat other psychiatric disorders, and even prisoners (Pearce & Haigh, 2017; Vandevelde et al, 2004).

The American case, however, draws a striking distinction. In the United States, TCs were developed with the intent of substance use disorder treatment in mind. For this application, the treatment emphasized peer influence and mutual assistance (De Leon, 2000; Clark, 2017). Residents in these "hierarchical" TCs are expected to monitor one another's behavior. Formally, the feedback provided between residents typically occurs through either written affirmations (praising a peer for their progress in treatment or social norms), or corrections (encouraging a peer to better adhere to their treatment or social norms) (De Leon, 2000; Hawkins & Wacker, 1986; Patenaude, 2004). While staff are expected to model appropriate behavior, senior residents play a significant role in serving as role models and mentors for more junior residents (De Leon, 2000).

Our major source of knowledge in defining and conceptualizing hierarchy within the TC context comes from De Leon (2000):

*"The peer hierarchy of work and community status defines the roles, functions, and relationships that mediate social and therapeutic change, while the peer culture, embodied in the norms, values, and beliefs of right living, guides the change process (151)."*

One important limitation, however, is noted:

*"However, since residents are learning how to manage themselves and the environment as well as how to interact with others, they are limited in their ability to guide each other in the change process (De Leon 2000, 152)."*

While staff may be the "ultimate authority in the clinical and community management" of the TC, residents are not powerless, and senior residents model and advise junior residents on community norms and treatment principles (De Leon 2000, 152). Given that we are often most concerned with how the mutual aid component of the TC model influences resident recovery, it seems necessary to further conceptualize what it means for a TC to be hierarchical. However, it is insufficient to just know what a hierarchy is, we must understand how to translate the concept to real world measurement in observational data. Only then can we systematically understand how the "hierarchical" orientation of a TC can influence residents' prospects for recovery. Specifically, there are a series of questions that we aim to address in the following sections: How should we define hierarchy within a TC context? How could we measure the hierarchy among TC residents from observational data? What factors tend to influence one's position in the hierarchy? What does one's position in the hierarchy reveal about their treatment? We explore these questions in this manuscript.

## Conceptualizing Hierarchy

Hierarchy permeates social systems, and researchers across the social sciences have dedicated attention to better understanding it. In political science, scholars study the hierarchical nature of international relations (Lake, 2011). In sociology, there has been an attempt to understand hierarchy within the context of academic hiring (Clauset et al., 2015). The work most heavily influencing ours comes from zoology (Hobson and DeDeo, 2015).

Any conceptualization of hierarchy should fulfill a few conditions. First, it must *allow* for some ranking of components such that some can be said to be at the "top" of the hierarchy, and some can be at the "bottom." This also establishes transitivity as an essential component of hierarchy. If A possesses some authority over B, and B possesses some authority over C, it necessarily implies that A possesses some authority over C. Rank and transitivity are well known, established components of hierarchical relations (Chase, 1980; Chase, 1982; Hobson and DeDeo, 2015; Hobson et al., 2018). While there is the potential for intransitivity in hierarchical relationships, what some call non-linear hierarchies (Chase, 1980; Chase, 1982), recent work has shown that components are relatively good at inferring and learning the broader hierarchy and most social systems conform to a linear hierarchy (De Vries, 1998; Hobson and DeDeo, 2015)

Second, it must allow for the volume of components to vary at each level of the hierarchy (Wellman et al., 2020). In some circumstances, a hierarchy might resemble a triangle or a pyramid, relatively few components at the top, and then more gradually until you reach the bottom of the hierarchy (Wellman et al., 2020). However, in other circumstances, it might resemble a diamond, having relatively few components at the top, the majority in the middle, and few in the bottom.

Third, hierarchy necessitates order. That order can be formal, like the reporting structure of a business (Wellman et al., 2020), or informal and emergent, like the hierarchy that emerges from dominance interactions between animals (Hobson and DeDeo, 2015). Order implies that there are a series of norms and expectations that guide action that are intrinsically tied to the ranking of components within a hierarchy. These norms and expectations influence how superordinates and subordinates interact with one another. For example, in the dominance hierarchies in the animal kingdom, superordinates typically have better access to foraging resources and reproductive opportunities (Dewsbury, 1982).

While these three components may not constitute the conceptual whole of hierarchy, they are the foundational components. They also are components found within what we would consider the "hierarchical" TCs common in the United States. In the following pages we will discuss this in greater detail, isolating not just what hierarchy looks like within the TC context, but how we can measure it using an extensive dataset of corrective interactions between TC residents.

## Operationalizing Hierarchy

Researchers relying upon observational data to study the hierarchy of a social system typically rely upon the use of directed social networks. These networks are comprised of a set of actors who relate to one another in some way; such relations can be symmetric/reciprocated or asymmetric/unreciprocated. Consider a friendship network. A may report being friends with B, but B may or may not report being friends with A. The directionality wherein one actor "sends" a relationship to another is what defines a network as either being directed or undirected. In a therapeutic community, residents have the opportunity to correct other residents. These corrections may accomplish a variety of clinical objectives, but they may also reveal information about one's status in the peer hierarchy (De Leon 2000, 151).

With a directed network that is negatively valenced such as this, eigenvector centrality is frequently used to measure hierarchy (De Vries, 1998; Brush et al., 2013). Conventionally, eigenvector centrality would be a measure of the influence an individual has within a social network. Eigenvector centrality measures not just how many social connections an individual has, but how well connected that individual's connections are. The intuition is that quality trumps quantity, and having a few well connected friends might be more valuable than having many friends (Bonacich, 1987). Whenever the network is negatively valenced, however, instead of implying influence and power, one's position in the pecking order is considered. In the example presented by Hobson and DeDeo (2015), Bird A might be pecked by B, who is then pecked by C and D. A would have the highest eigenvector centrality as they are pecked by a bird who is of less status, being pecked by C and D. In our study, being corrected by a resident who is themselves corrected by many would imply being of lower status.

Eigenvector centrality has been used to measure hierarchy in a variety of academic disciplines and contexts with documented success. In zoology, Hobson, Mønster, and DeDeo (2018) use eigenvector centrality to study the hierarchies present in 172 social groups across 85 species in 23 distinct orders. In sociology, there have been many studies that have used measures like eigenvector centrality to measure prestige and hierarchy across universities or academic departments (Myers et al., 2011; Clauset et al., 2015; Fowler et al., 2007). Beyond these interesting applications, eigenvector centrality has been used to measure hierarchy in studies of brain networks (van Duinkerken et al., 2017), terrorist networks (Memon and Larsen, 2006), and judicial precedents (Hitt 2016).

Eigenvector centrality captures the previously discussed, necessary components of our conceptualization of hierarchy. First, it is a continuous measure bound [0,1] that captures the gradient of highest position in the hierarchy (marked by low values, like zero) to the lowest position in the hierarchy (marked by high values, like 1). This allows us to get a full sense of the rank of residents within the TC hierarchy. Second, the only thing that influences the "shape" of the hierarchy (e.g., triangle, diamond) is the composition of the network. In other words, eigenvector centrality allows us to capture the shape of the hierarchy without any *a priori* assumptions or external constraints, allowing the data to speak for itself. Third, while eigenvector centrality cannot tell us the norms or expectations stemming from the order emerging from a hierarchy, it allows us to encode assumptions about the action stemming from these expectations in the data. For example, when using data on corrections, we can "back into" the hierarchy by using knowledge about who corrects whom, knowing that residents use corrections to mentor others in the TC.

When constructing this longitudinal network of corrections, we use previously collected data of resident-to-resident corrections from a large TC unit from a large Midwestern state. We compose a set of longitudinal weekly social networks for each unit measuring the number of times residents corrected one another. With this operationalization of hierarchy in mind, we can gain a better understanding of how our measurement comports to our prior expectations and provide some firm evidence for clinical conventional wisdom.

## Results

To better understand the hierarchy that structures interactions within a TC, we ask and answer three different sets of questions using data on corrections between residents at a large facility in a large Midwestern state. Specifically, we shed light on the following:

- What is the shape of the hierarchy in the most active week during observation? When the hierarchy was at its peak, was it a triangle or more of a diamond?
- As a resident spends more time in a facility and they become more "senior", is this reflected in the hierarchy? Do residents tend to have higher positions in the hierarchy at the end of their tenure, relative to the beginning of their tenure?
- Does position in the hierarchy, which may reflect an internalization of TC principles, correspond to a higher rate of graduation?

## Shape of the Hierarchy

The shape of the hierarchy within a TC is an important topic that better allows us to understand the means through which mutual peer aid works. Is the hierarchy triangular, wherein there are relatively few residents at the top and many at the bottom? Is that triangle inverted, with many residents at the top and few at the bottom? Is it a diamond, with few residents at the top and bottom, but many in the middle? If we understand the shape, or orientation of this hierarchy we can have a better understanding of how peer influence works in the aggregate. Each shape would imply a different set of dynamics.

To examine the shape of the hierarchy within our TC of interest, we examine the network of peer-to-peer corrections during the week with the most activity. We choose to focus on the most active week as it should, in principle, be the most visible manifestation of the hierarchy. During this week, we observe 82 residents correcting one another 322 times. This network is presented in Figure 1, where residents are black dots connected by corrections (grey lines). The residents are sized by their inverse eigenvector centrality with those at the highest end of the hierarchy having the largest dots.

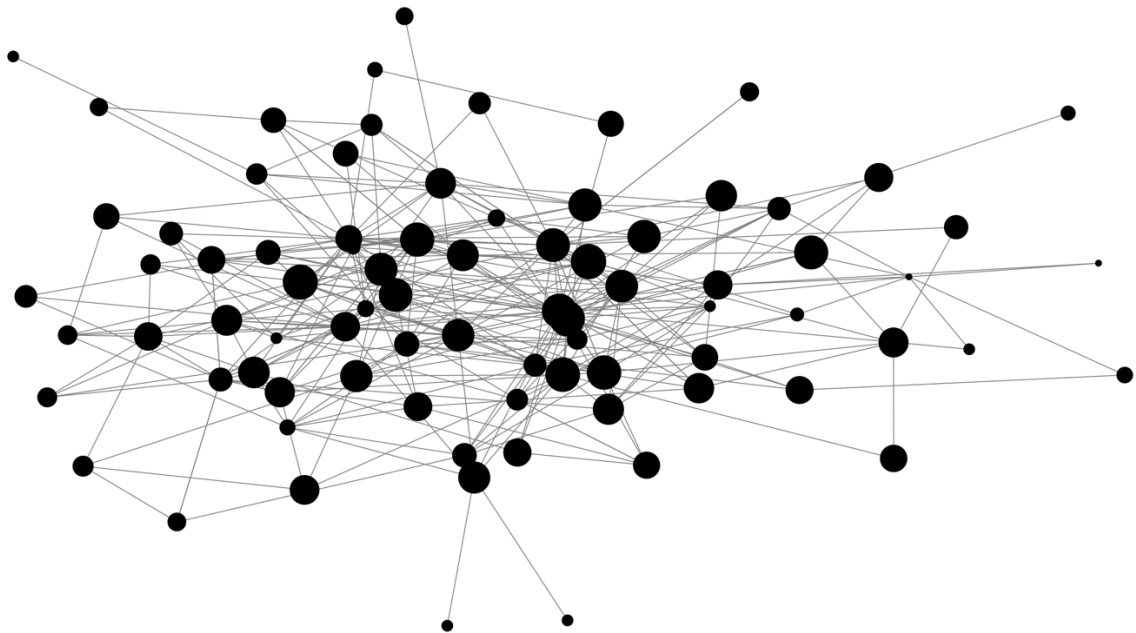

Figure 1: Network Plot, Nodes Sized by Inverse Eigenvector Centrality

Overall, we find that the hierarchy for this TC appears to be something of a top-heavy diamond. There are a handful at the top, many just below the top, and then relatively few at the bottom. This is shown by the histogram in Figure 2. This plot shows the relative frequency of certain eigenvector centrality values with their density overlaid. You'll see that there are a number of residents who achieve the highest position in the hierarchy (marked by a value of 0.00), then more as you move from the top of the hierarchy to the bottom, only to start a continuous decline after an eigenvector centrality of around 0.12.

The left-skewedness of this distribution, paired with a shallow but long-tail highlights three features of clinical importance. First, the vast majority of residents tend to do relatively well, achieving higher positions in the hierarchy. Second, because the vast majority of residents seem to do relatively well, one might think that the facility and the hierarchy is robust to residents graduating on from the program and leaving as another individual could fill in. Third and finally, there are relatively few who regardless appear to do very poorly and have been highlighted by the hierarchy as needing additional support. This is something of a negative reputation effect, after residents receive a correction there appears to be an increased likelihood of receiving another, potentially from a distinct peer, in the short-term. This may be a useful mechanism as it directs feedback where it is most needed.

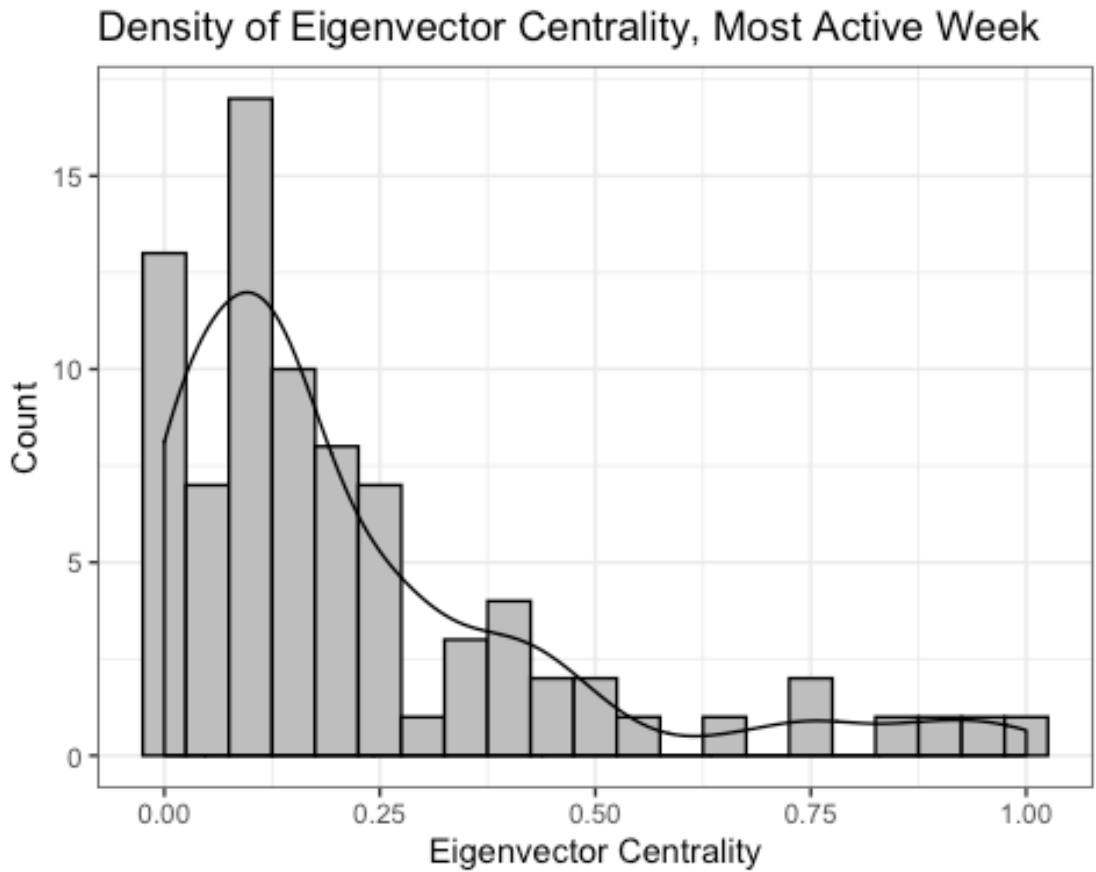

Figure 2: Histogram of Eigenvector Centrality for Most Active Week, Density Overlaid

This reveals meaningful information about the structure of a TC's hierarchy. While there may be a handful of residents on the top, they advise the bulk majority of those in the facility, with the majority advising the remaining residents towards the bottom of the hierarchy.

Hierarchy and Seniority

Clinical conventional wisdom would hold that as residents progress through their treatment program, gaining seniority, their position in the hierarchy should increase. In other words, we would expect that a resident's terminal position in the hierarchy should be higher than their starting position in the hierarchy. The thinking is that as residents become more senior, they better internalize the norms associated with the facility, make better progress on their treatment program, and become role models to new residents.

To examine the average trajectory of a resident in the facility, we present Figure 3. Figure 3 highlights the global average of individualized 4 week rolling averages of eigenvector centrality. In other words, we calculate a 4 week rolling average for each individual across their time in the program, and then take the average of those at each time period. For example, the point corresponding to week 10 would be the average of each resident's rolling average during their

10th week in the program. What you'll notice is that during the first few months in the program, individuals tend to have a consistent middle-of-the-pack eigenvector centrality which starts to pick up during the middle of their program, only to drop-off towards the end of their program. This bump in the middle makes sense as it should reflect when residents start to hit their stride and likely start being held more accountable than in the start of their program. This finding carries with it some important clinical guidance, clinicians and peers should reassure residents that things might look like they're getting worse before they're getting better, but in reality, important progress is being made which might not be visible until the end of their program. We notice that one's position in the hierarchy starts to increase rapidly after the 20th week in the program (their penultimate monthy). This indicates that on average towards the end of their program, residents tend to be in a good position and naturally have a higher position in the hierarchy as a result of their clinical progress.

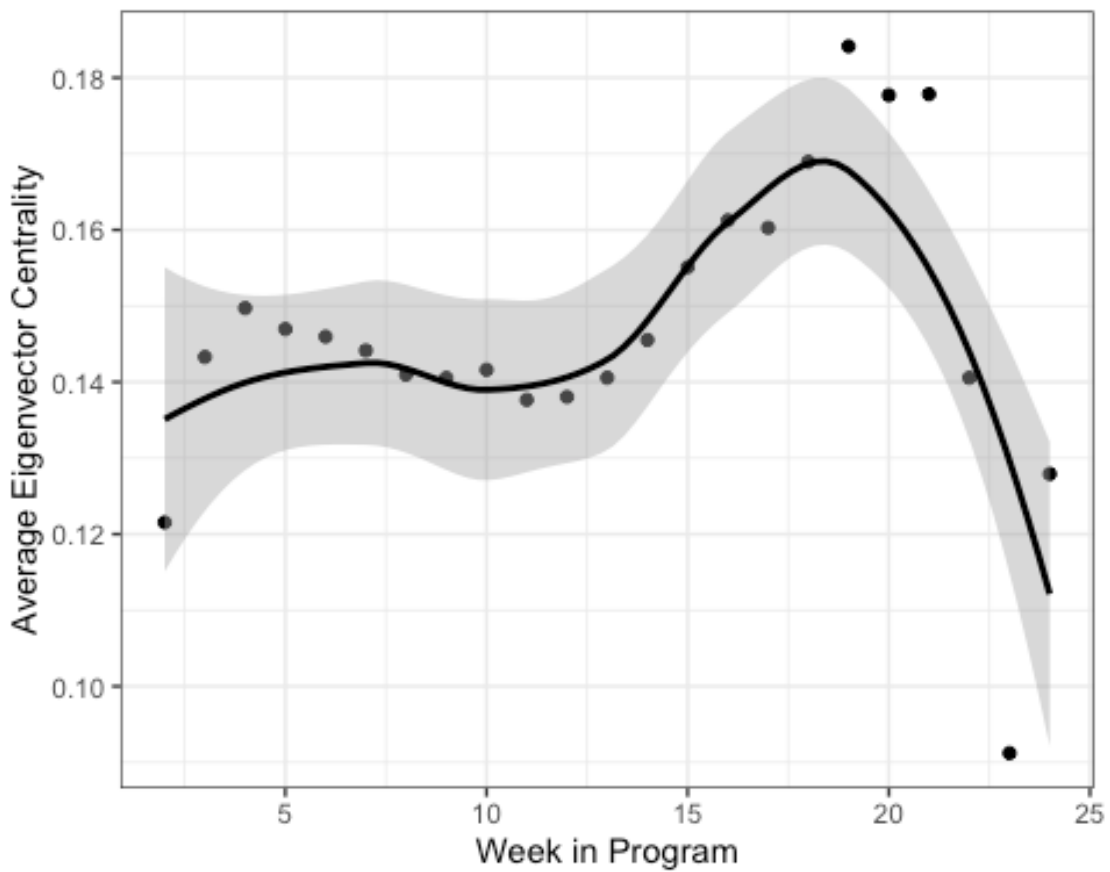

Figure 3: Global Average of Individualized Four Week Rolling Averages of Eigenvector Centrality for Standard Term of Program (24 Weeks)

To test this proposition formally, we calculate the average eigenvector centrality for the first 4 weeks and last 4 weeks of a resident's treatment program. We present these values in Figure 4, which shows the boxplots for the two measures side-by-side. Overall, you'll notice that there are some noticeable differences in the distributions of these variables, although the contrast may not be as stark as we might otherwise expect given the long tail of each distribution.

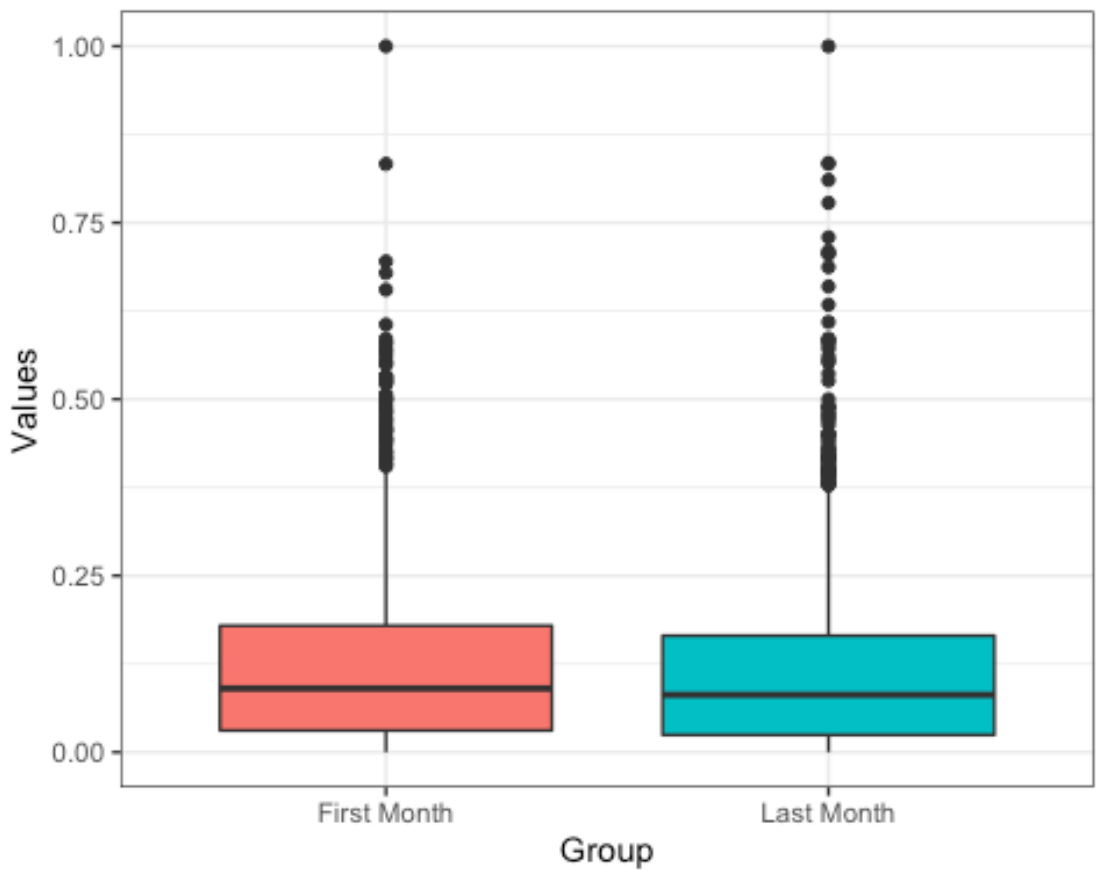

Figure 4: Boxplots for Average Eigenvector Centrality

To make better sense of these differences and to determine whether one's terminal position in the hierarchy is higher than their starting position, we use two statistical tests that capture distinct dynamics. First, we use Wilcoxon paired tests which will allow us to test for differences between these starting and ending values in a paired, resident-level way. Second, we use Kolmogorov-Smirnov tests to compare the cumulative distribution functions for the two sets of values. This allows us to contrast the data generating processes to determine if they are statistically distinct.

Both tests confirm our theoretical expectations. For the Wilcoxon paired tests, we reject the null hypothesis that the value change between the average eigenvector centrality for the first and last month is equal to zero ($p < 0.05$). Given the directionality implied in Figure 4, this leads us to conclude that there are meaningful improvements in one's position in the hierarchy as they become more senior. For the Kolmogorov-Smirnov test, we reject the null hypothesis that the variables were drawn from the same data generating processes ($p < 0.05$). This too leads us to believe that in aggregate, these variables are drawn from distinct data generating processes.

This evidence indicates that clinical conventional wisdom isn't too far off from what we are seeing in our data – there are important, statistically meaningful differences between where a resident starts in the hierarchy, and the position they reach as they become more senior.

### Hierarchy and Graduation

The final question we examine is whether one's terminal position in the TC hierarchy predicts whether they successfully graduate from their treatment program. It is thought that one's position in the hierarchy should track with their internalization of TC norms and success in adhering to their treatment program. In fact, the entire mutual aid, peer influence dynamics of the TC treatment model depend upon this proposition being true. If the residents who are influencing the behavior of others are not those that should serve as role models, then the socialization benefits of the TC model would be undermined. Alternatively, if those who should be serving as role models are not, then the socialization benefits of the TC model would be undermined.

To answer this question, we examine the bivariate relationship between one's average eigenvector centrality during the last four weeks of their treatment program with whether they successfully graduated from the program. It is worth noting that there is some variance in the outcome; while the majority of residents do complete their treatment program, 17% drop out. While there are a variety of other factors that may influence one's graduation, we save a more rigorous study of this question for future work.

In a bivariate logistic regression model, we find that those with a higher average eigenvector centrality over their last month (implying a lower relative position in the hierarchy) are less likely to graduate. This relationship ($\beta = -4.91, \sigma = 0.49$) is significant at any conventional threshold ($p < 0.001$). A visualization of this relationship is presented in Figure 5. The baseline probabilistic expectation that someone with the highest position in the hierarchy graduates is around 0.88, which drops all the way down to less than 0.10 for someone with the lowest position in the hierarchy. This indicates that there is a substantively meaningful relationship between these two factors in the data.

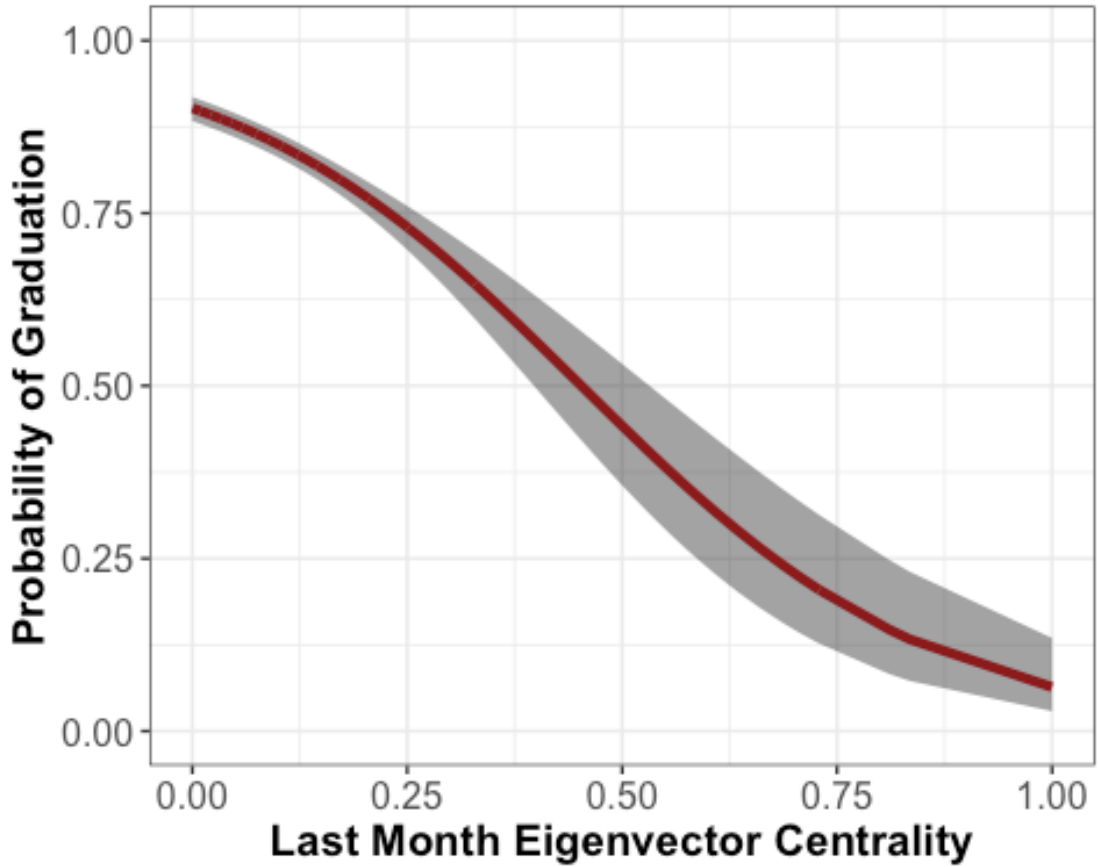

Figure 5: Predicted Probability Curve for Eigenvector Centrality-Graduation Logistic Regression

## Conclusion

In this piece, we have introduced a framework that allows us to better understand and study the hierarchies that constitute the TC clinical model for substance use disorder recovery. We applied this new framework to answer a series of questions that are of importance to TC researchers and clinicians. The framework introduced in this piece, provides support for three pieces of clinical conventional wisdom. First, we find that the hierarchy within the TC resembled something of a diamond, with a handful of residents at the top of the hierarchy, more in the middle, and a handful at the bottom. This allows us to understand the aggregate mentorship dynamics that underlie the peer influence, mutual aid model. Second, we find that a resident's terminal position in the hierarchy is typically higher than their initial position, indicating that as a resident becomes more senior, they become more of a mentor to other residents. Finally, we find that the higher one's terminal position in the hierarchy, the more likely they are to successfully complete their treatment program. This implies that those who mentor peers are those who should be mentoring others as they have made progress in their treatment program and understand the norms of the community.

At least two limitations should be mentioned. The first is that this is a study of a limited sample and that there is no guarantee of external validity. The second is that the analysis as it stands does not control for such possible confounders as race, gender, or age. While these are legitimate caveats, the theory behind the analysis has seen application in previous studies of other complex systems, so it is reasonable to think that a similar hierarchy might be found in other TCs.